\documentclass[fleqn,usenatbib]{mnras}

\usepackage{newtxtext,newtxmath}
\usepackage[T1]{fontenc}

\DeclareRobustCommand{\VAN}[3]{#2}
\let\VANthebibliography\thebibliography
\def\thebibliography{\DeclareRobustCommand{\VAN}[3]{##3}\VANthebibliography}

\usepackage{graphicx}	% Including figure files
\usepackage{amsmath}	% Advanced maths commands
\usepackage{orcidlink}

\newcommand{\refnote}[1]{#1}

\title[CETRA]{CETRA: A fast, sensitive exoplanet transit detection algorithm implemented for GPUs}

\author[L. C. Smith et al.]{
Leigh C. Smith\orcidlink{0000-0002-3259-2771},\thanks{lsmith@ast.cam.ac.uk}
Saad Ahmed\orcidlink{0000-0001-8304-5586}, 
Francesca De Angeli\orcidlink{0000-0003-1879-0488},
P. W. Burgess\orcidlink{0009-0002-6668-4559},
Giorgia Busso\orcidlink{0000-0003-0937-9849},
\newauthor
Dominic C. Ford,
Diana L. Harrison\orcidlink{0000-0001-8687-6588},
S.~T.~Hodgkin\orcidlink{0000-0002-5470-3962},
Jonathan M. Irwin,
Guy T. Rixon\orcidlink{0000-0003-4399-6568}, 
\newauthor
and Nicholas A. Walton\orcidlink{0000-0003-3983-8778}
\\
Institute of Astronomy, University of Cambridge, Madingley Rd, Cambridge CB3 0HA, UK
}

\date{Accepted 2025 March 26. Received 2025 March 26; in original form 2025 March 13}

\pubyear{\the\year{}}

\begin{document}
\label{firstpage}
\pagerange{\pageref{firstpage}--\pageref{lastpage}}
\maketitle

\begin{abstract}
We present the Cambridge Exoplanet Transit Recovery Algorithm (\textsc{cetra}), a fast and sensitive transit detection algorithm, optimised for GPUs. \textsc{cetra} separates the task into a \refnote{search for} transit \refnote{signals across linear time space,} followed by a phase-folding of the former \refnote{to enable} a periodic signal search, using a physically motivated transit model to improve detection sensitivity. It outperforms traditional methods like Box Least Squares and Transit Least Squares in both sensitivity and speed. Tests on synthetic light curves demonstrate that \textsc{cetra} can identify at least 20 per cent more low-SNR transits than Transit Least Squares in the same data, particularly those of long period planets. It is also shown to be up to a few orders of magnitude faster for high cadence light curves, enabling rapid large-scale searches.

Through application of \textsc{cetra} to Transiting Exoplanet Survey Satellite short cadence data, we recover the three planets in the \mbox{HD 101581} system with improved significance. In particular, the transit signal of the previously unvalidated planet \mbox{TOI-6276.03} is enhanced from ${\rm SNR}=7.9$ to ${\rm SNR}=16.0$, which means it may now meet the criteria for statistical validation.

\textsc{Cetra}’s speed and sensitivity make it well-suited for current and future exoplanet surveys, particularly in the search for Earth analogues. \refnote{Our implementation of this algorithm uses NVIDIA’s CUDA platform and requires an NVIDIA GPU,} it is open-source and available from GitHub and PyPI.
\end{abstract}

\begin{keywords}
exoplanets -- stars: planetary systems -- software: data analysis
\end{keywords}

%%%%%%%%%%%%%%%%%%%%%%%%%%%%%%%%%%%%%%%%%%%%%%%%%%

%%%%%%%%%%%%%%%%% BODY OF PAPER %%%%%%%%%%%%%%%%%%

\section{Introduction}

Three quarters of the current list of several thousand confirmed exoplanets were discovered through the observation of a reduction in the apparent flux of their host star as it is transited by the planet. %4\,210 out of 5\,671 at the time of writing, according to the NASA Exoplanet Archive\footnote{\url{https://exoplanetarchive.ipac.caltech.edu}}. 
The retired Kepler space telescope is responsible for the majority of these discoveries, through the Kepler and K2 missions \citep{kepler,K2}. The Transiting Exoplanet Survey Satellite \citep[TESS,][]{TESS} is responsible for several hundred of them, but it has generated many times more candidates that are awaiting independent confirmation. The upcoming PLAnetary Transits and Oscillations of stars \citep[PLATO,][]{PLATO} mission, with a planned 2026 launch, is designed specifically to detect and characterise terrestrial planets around Sun-like stars.

The depth of an exoplanet transit is approximately proportional to the square of the ratio of the planet and star radii. The fractional reduction in flux from a Sun-like star due to the transit of an Earth-like planet is of the order $10^{-4}$, or $100$~parts-per-million (ppm). This level of photometric precision is essentially unachievable from the ground, but measurements made by space-based instruments highlight a further problem: the amplitude of the intrinsic variability of the star is typically many times larger than such a transit signal \citep{kepler_variability}. Many large-scale exoplanet transit detection pipelines rely on a pre-whitening filter, i.e. they attempting to remove correlated noise (intrinsic stellar variability but often also instrumental artifacts) such that only the transit signal and uncorrelated (white) noise remain, before running a transit detector. For targets whose characteristic variability is particularly challenging for transit detection (amplitude and/or frequency-wise) it can be beneficial to attempt a simultaneous fit of the correlated noise \citep{nuance}. However, this approach is relatively expensive computationally, and doubt has been cast on the need for it in most situations \citep{sim_detrend_doubt}.

The standard transit detection algorithm has for many years been the Box-fitting Least Squares algorithm (BLS) of \citet{BLS}. This approach phase-folds the light curve over a grid of periods and finds the least-squares box-shaped signal for each using a moving window approach. \citet{TLS} demonstrated that the sensitivity of this type of detector could be improved upon by using a transit-shaped profile, which is particularly important for low signal-to-noise ratio (SNR) transits, and they presented their Transit Least Squares (TLS) algorithm. Even modest improvements to sensitivity at low SNR are valuable since this is where prized Earth-Sun analogues typically reside. Despite the additional computational complexity over BLS, TLS is not significantly more expensive as a result of optimisations made to the period and duration search grids.

Another design choice is whether to search for periodic signals in phase-folded data, or to search for signals in linear \refnote{time} data and phase-fold the results. Both BLS and TLS use the former approach, which requires an expensive sorting step in phase space for each trial period. If many light curves with the same observation times are available, one might perform sorting once per period for all and share the sorted indices. This optimisation is used by e.g. \citet{ppp}. The linear-then-periodic approach offers reduced computational expense but tends to be used primarily in algorithms that incorporate simultaneous detrending (e.g. \citealp{Mearth_tda}, \citealp{DFM15}, \citealp{nuance}; with the notable exception of the one used in the Kepler pipeline and described by \citealp{kepler_transits}). A preliminary search in linear space also enables a search in these results for single transits (a.k.a. monotransits), which are early indicators of the presence of long period planets.

The increasing volumes of data being supplied by current and future space-based instruments \refnote{encourage} the use of highly efficient transit search algorithms. Graphics Processing Units (GPUs) have potential here, since detection algorithms typically comprise highly parallel operations, operations which GPUs are ideally suited to perform.

We present a new transit search algorithm, the Cambridge Exoplanet Transit Recovery Algorithm (\textsc{cetra}). \refnote{It performs the majority of the processing on} NVIDIA GPUs using the CUDA parallel computing platform, \refnote{and hence requires a compatible device}. \textsc{Cetra} incorporates the ideas behind several existing algorithms, while also including additional optimisations in order to try to improve on their sensitivity and/or compute efficiency. It is designed to be run on detrended light curves, uses an exoplanet transit shaped model, and is a two-stage linear-then-periodic algorithm. In this article, we detail the design and implementation of \textsc{cetra} (Section \ref{sec:algo_description}), and evaluate its computational performance and detection sensitivity relative to established algorithms (Section \ref{sec:performance}). We then reanalyse the \mbox{HD 101581} multi-planet system in the TESS high cadence data, resulting in an increased transit signal detection significance \refnote{relative to that of the existing literature} (Section \ref{sec:HD101581}).
\textsc{Cetra} is published on github\footnote{\url{https://github.com/leigh2/cetra}} and PyPI under an open-source MIT license.

\begin{figure}
  \begin{center}
    \includegraphics[width=0.45\textwidth,keepaspectratio]{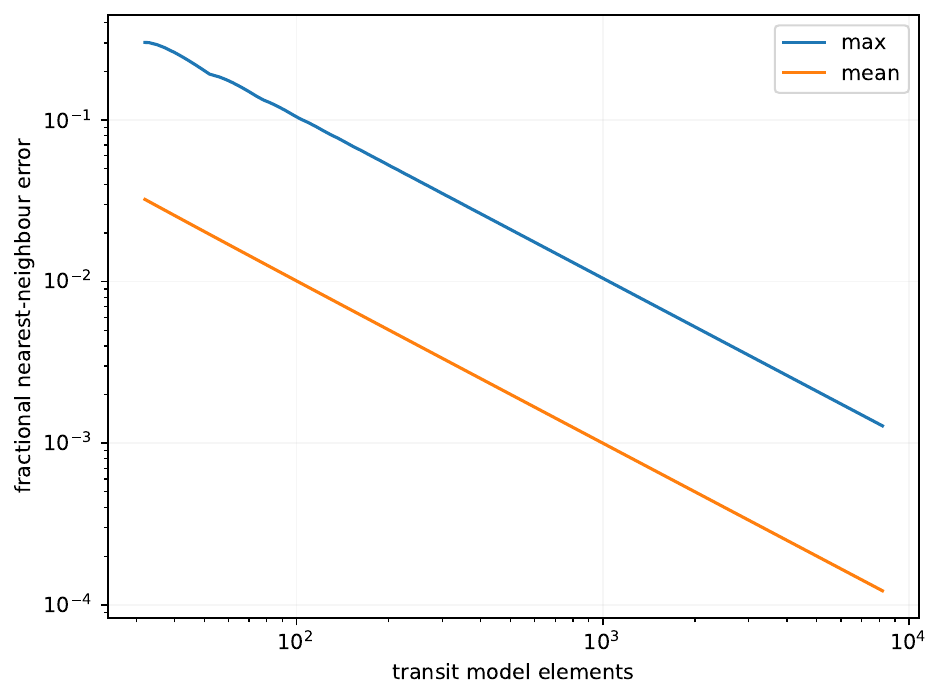}
    \caption{The maximum and mean error introduced by the nearest neighbour \refnote{transit} model point selection method used by the linear search kernel, for \refnote{an example exoplanet transit model with a range of lengths (x-axis). The length of the exoplanet transit model (in array elements) is the number of samples of the input transit model after the resampling stage.} This can be used to inform the selection of a suitable transit model resampling cadence.}
    \label{fig:tm_error}
  \end{center}
\end{figure}

\section{Algorithm Description}
\label{sec:algo_description}

\textsc{Cetra} searches light curves for transit-like features using a true transit model, since this offers improved sensitivity relative to a box model, but with detection separated into linear and then periodic signal searches. The linear \refnote{search} component finds the maximum likelihood model depths \refnote{over a two dimensional grid, with trial durations on one dimension and an array of $t_0$s that is linear in time on the other}. The periodic \refnote{search} component then phase-folds the \refnote{output} from the linear search over a grid of periods and returns the maximum joint-likelihood for each. \textsc{Cetra} assumes that light curves it is given have already been detrended. More detail about each component, and also about some important setup stages, is provided below.

Internally, \textsc{cetra} works in units of normalised flux offset from baseline, i.e. baseline flux is `$0.0$' and \refnote{if flux from the star were blocked completely then its normalised flux offset would be} `$1.0$'. Time units are days, and where appropriate, those relative to the light curve start time are used. These conventions mean that the use of single precision floating point numbers introduces smaller errors than otherwise. In the case of the flux for example, using a baseline of `$1.0$` introduces a constant single precision floating point error of the order $0.1~{\rm ppm}$, where using `$0.0$` incurs an error that is proportional to the flux deviation and only reaches that level when no flux is observed. For times, using this convention incurs a floating point error of only a few seconds over light curves that are several years long. These are acceptable for the purposes of transit detection, and the speed difference between single and double precision operations can be significant on some GPU models. However, these conventions are broadly transparent to the user.

\subsection{Setup}
\label{sec:algo_setup}
\subsubsection{Light curve resampling}\label{sec:algo_resample}

The first setup stage is a resampling of the input light curve into a regular, ordered time grid. For example, PLATO Level 1 light curve points are averages between camera groups \citep{PLATO}, which leads to an irregular cadence. Even a strictly regular on-board cadence will yield an irregular cadence when converted to Barycentric Dynamical Time. The main advantage of the resampling is that the resampled light curve array element corresponding to any point in time can be determined with ease. For example, in the context of an exoplanet transit, the first and last resampled light curve element indices of the window covering the transit are simply 
\mbox{$\lceil{}t / {\it cadence}\rceil{}$} 
and 
\mbox{$\lfloor{}(t + {\it duration}) / {\it cadence}\rfloor{}$},
respectively, where $t$ is the start time of the transit relative to the start time of the resampled light curve, \textit{duration} is the duration of the transit, and \textit{cadence} is the cadence of the resampled light curve. Resampling in this way avoids any need for repeated, expensive sorting of the light curve, and makes the algorithm robust against gaps and irregular cadence in the input data.

The frequency of points on the resampled grid is a choice for the user. Observed fluxes are simply assigned to the nearest output point. Where multiple input observations contribute to single output points they are combined using an inverse variance weighted mean and their uncertainties are propagated formally. Resampled grid points that have no associated observational data are assigned \textit{null} fluxes with infinite error. If no resample cadence is supplied by the user then \textsc{cetra} will use the median cadence of the input data by default.

It is generally preferable to resample to a higher frequency than the input observations, since this limits morphological distortions of the light curve. \citet{dont_bin} demonstrates the potential impact of distortions in the context of binning. In terms of computational expense, the frequency of the resampled light curve only impacts the linear component of the transit search, and we demonstrate in Section \ref{sec:performance_compute} that a relatively high frequency is not particularly costly.

An additional detail is that the resampled light curve is also padded with null data at the beginning and end, with data length equal to half the maximum trial duration (see Section \ref{sec:algo_grids}). This is a straightforward way to ensure that the linear search can detect transits that straddle the beginning and end of the data.

\subsubsection{Transit model}\label{sec:algo_tmodel}

\textsc{Cetra} includes densely sampled transit light curves produced using the \textsc{batman} package\footnote{\url{lkreidberg.github.io/batman}} \citep{batman} for three different values of impact parameter. The provided models use the following parameters:
\begin{description}
    \item Planet radius $R_p=0.03R_*$,
    \item impact parameter $b=[0.32, 0.93, 0.99]$,
    \item and quadratic limb darkening parameters $u_1,u_2=0.4804,0.1867$.
\end{description}
These are also the default model parameters used by TLS (in the $b=0.32$ case). However, users might wish to supply their own models instead, since the default parameters are not ideal for all targets. For example, if the limb darkening parameters of the target are known, even approximately e.g. from atmospheric modelling, then a model that uses these might be supplied instead and may improve detection sensitivity. Additionally, since changes to the impact parameter can make a significant difference to the shape of the model, it may be worth running \textsc{cetra} across a sequence of values. An optimal set of parameters is likely to depend on survey design and sensitivity requirements, and is outside the scope of this paper.

%User supplied models should be in the form of a 1D array of flux offsets from baseline. They should be normalised by the maximum offset from baseline, the first point should be the start of the transit, and the model should be no longer than the transit itself. \textsc{Cetra} outputs transit depths, start times, and durations under the assumption that this is the case. If it is not the case then additional care must be taken when interpreting its outputs.

The chosen/supplied model is ultimately resampled to a user-defined number of samples using simple linear interpolation. This is done to ensure that the model will fit into shared memory on the GPU alongside some additional arrays. The linear search kernel\footnote{A GPU function is a \textit{kernel}.}
makes frequent comparisons of the light curve to the model, and hence the much lower latency of shared memory makes its use worthwhile. 

\refnote{Within} the linear search kernel the nearest model point is selected for a given light curve point, transit reference time and duration being considered. This nearest neighbour selection introduces a small error in flux, which is greatest where the flux gradient is largest. The maximum and mean error introduced versus the number of samples chosen is shown in Figure \ref{fig:tm_error} for the default transit model. The maximum error occurs during ingress and egress, but the mean error is around an order of magnitude smaller. Since the volume of shared memory available is limited, and memory copies should be be reduced for efficiency where practical, it is advisable to use as few model samples as necessary to achieve an acceptable level of error. For reference, the linear search requires 256 bytes of shared memory space in addition to the transit model. The default number of samples is $1024$, for which the maximum error is $\approx{}1$~per~cent and the mean error is $\approx{}0.1$~per~cent.

It is worth noting that while \textsc{cetra} was written with exoplanet transit searches in mind, any model can be supplied to it (e.g. exocomets, exoplanets with rings, etc.). In principle it might be used for any 1D template matching purpose, unrelated to astronomical light curves or even to astronomy, though some modification might be necessary. No assumption is made about whether the deviation of the model is positive or negative, it does not enforce a minimum depth, so e.g. stellar flares and other such flux-positive transient phenomena might also be searched for instead. For alternative use cases consideration must be made as to whether the default duration (and period) grids are appropriate (see Section \ref{sec:algo_grids}).% In this paper we generally refer to exoplanet transits for examples, but any other model is typically also applicable.

\subsubsection{Duration, t$_0$ and period grids}\label{sec:algo_grids}

The linear transit search is performed over grids of transit reference times, $\{t_{i}\}_{i=1}^{I}$ with $I$ times, which are the mid-point of the transit; and durations, $\{D_{j}\}_{j=1}^{J}$ with $J$ durations. Subsequent periodic signal searches are performed over a grid of periods, $\{P_{k}\}_{k=1}^{K}$ with $K$ periods.

By default, \textsc{cetra} uses the same TLS specification for the duration and period grids, although the user is free to supply their own of either. For a given period, durations that fall outside those expected for typical Keplerian orbits (see figure 5 of \citealt{TLS} in particular) are not evaluated, though the user can override this behaviour. A major optimisation of TLS is the use of duration and period grids that are astrophysically motivated and optimally sampled \citep{TLS,ofir14}. We saw no reason to deviate from these grid definitions.

The \textsc{cetra} reference time grid uses a single step size for all durations. The default is 1~per~cent of the minimum duration, although the user is free to specify an alternative fraction of the minimum duration. TLS uses a transit reference time grid that is a function of the trial duration, which means that there are fewer trial reference times for longer durations. We might adopt this optimisation for \textsc{cetra}. However, the fixed time step simplifies the algorithm and its output, and means that the results of the linear search could be directly probed for transit timing variations \citep[TTVs; adopting the approaches of e.g.][]{QATS,RIVERS} without a reduction in sensitivity for long-duration transits. The density of the reference time grid impacts the compute time of the linear search only, and as is demonstrated in Section \ref{sec:performance_compute}, increasing the run time of the linear search is not a major burden.

\subsection{Linear search}
\label{sec:algo_linear}

The linear search traverses the grid of transit reference time, $\{t_{i}\}_{i=1}^{I}$, and transit duration, $\{D_{j}\}_{j=1}^{J}$, element pairs. Each pair defines a region in time in which the light curve should be compared to the transit model, this time span will hereafter be referred to as the transit window. Because \textsc{cetra} resampled the light curve during setup, it can trivially determine the section of data that covers the transit window, see Section \ref{sec:algo_resample}. This means that expensive whole-array traversals are not required, and only two traversals of the subset of the light curve array corresponding to the transit window are needed. The first pass computes the maximum-likelihood transit depth, by which the transit model fluxes must be multiplied in order to optimally match the observed fluxes. The second pass computes the log-likelihood of the observed fluxes given the scaled model fluxes, and subtracts from this the log-likelihood of a constant flux model to obtain the likelihood ratio between the two.

With a transit model normalised to maximum depth $=1$, obtaining an equivalent model with a transit depth of $\Delta{}$ is simply a matter of multiplying each element by a scale factor of $\Delta{}$. In a given transit window, the model scale factor (i.e. transit depth) implied by each flux element, $f_n$, is $\Delta{}_n = f_n/m_n$, where $m_n$ is the appropriate model element obtained using nearest neighbour interpolation (see Section \ref{sec:algo_tmodel}). The uncertainty on the model scale factor is $\sigma{}_{\Delta{}_n} = \sigma{}_{n}/m_n$, where $\sigma{}_{n}$ is the uncertainty on the flux. The maximum likelihood scaling factor, $\Delta{}_{i,j}$, over the window is the inverse variance weighted mean of all values of $\Delta{}_n$ inside the window, and the variance on this, $\sigma_{\Delta{}_{i,j}}^2$, is the reciprocal of the sum of their weights. These two parameters give us the maximum likelihood transit depth within the window and its error, assuming the model is appropriate. However, determination of the likelihood of the model requires a second traversal.

Recall that \textsc{cetra} works in normalised flux offsets from baseline internally, and that input light curves will have been detrended. As a result, out of transit model fluxes are zero.

For a vector of fluxes $\mathbf{f}$, i.e. a light curve of length $N$, the log-likelihood of the $t_{i}$,$D_{j}$ model is:
\begin{equation}\label{eqn:logl}
    \ln{}p(\mathbf{f}|t_{i},D_j,\Delta{}_{i,j}) = \sum_{n=1}^{N}\biggl(
    -\frac{1}{2} \frac{(\Delta{}_{i,j}m_n - f_n)^2}{\sigma{}_{n}^2} - \frac{1}{2}\ln{(2\pi{}\sigma{}_{n}^2)}\biggr)
\end{equation}
Computing the result of Equation \ref{eqn:logl} explicitly would require traversal of the entire light curve. However, the contribution of all out of transit points doesn't change between a transiting planet model and a constant flux model (i.e. a model containing no transit). If instead of Equation \ref{eqn:logl} we compute the likelihood ratio ($LR$) between these two models we only need to consider points within the transit window. As an added bonus, the second term inside the sum also cancels inside the transit window, so the $\log$ operation is avoided. The likelihood ratio, given in Equation \ref{eqn:lrat}, can therefore be computed at a much reduced cost relative to the full log-likelihood.

\begin{figure*}
  \begin{center}
    \includegraphics[width=\textwidth,keepaspectratio]{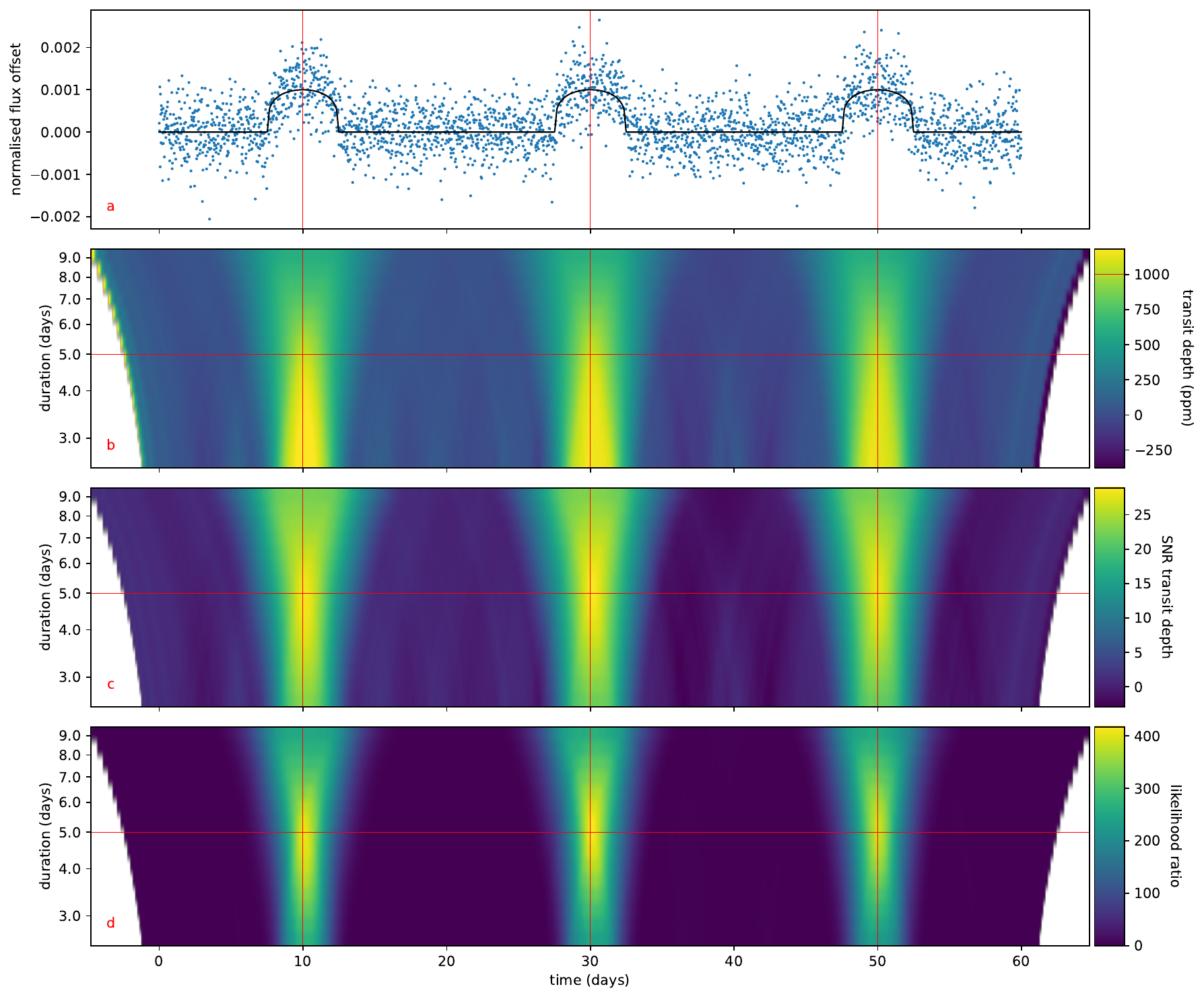}
    \caption{The output of the linear transit search for an example transiting planet light curve. The true values of $t_0$ and duration are indicated by red lines in the main panels, and depth by the red line in the colour bar of panel (b). Panel (a) shows the (astrophysically implausible) example input transiting planet model (black line) and associated light curve (blue points); panel (b) shows the maximum likelihood transit depth; panel (c) shows the SNR of the maximum likelihood transit (the depth divided by its error); and panel (d) shows the likelihood ratio of the light curve given the maximum likelihood transit model versus a model containing no transits.}
    \label{fig:linear_output}
  \end{center}
\end{figure*}

\begin{equation}\label{eqn:lrat}
\begin{split}
    LR_{i,j} &= \frac{p(\mathbf{f}|t_i,D_j,\Delta{}_{i,j})}{p(\mathbf{f}|f=0)}\\
    & = \ln{p(\mathbf{f}|t_i,D_j,\Delta{}_{i,j})} - \ln{p(\mathbf{f}|f=0)} \\
    %& = \sum_{i=t_s}^{t_s+D}\biggl( -\frac{1}{2} \frac{(\Delta{}m_i - f_i)^2}{\sigma{}_{i}^2} - \frac{1}{2}\ln{(2\pi{}\sigma{}_{i}^2)}\biggr) \\
    %& - \sum_{i=t_s}^{t_s+D}\biggl(-\frac{1}{2} \frac{f_i^2}{\sigma{}_{i}^2} - \frac{1}{2}\ln{(2\pi{}\sigma{}_{i}^2)}\biggr) \\
    & = \sum_{n=t_i}^{t_i+D_j}\biggl(
    -\frac{1}{2} \frac{(\Delta{}_{i,j}m_n - f_n)^2}{\sigma{}_{n}^2}\biggr) -\sum_{n=t_i}^{t_i+D_j}\biggl(-\frac{1}{2} \frac{f_n^2}{\sigma{}_{n}^2}\biggr) %\\
    %& = \sum_{i=t_s}^{t_s+D}\biggl(-\frac{1}{2} \frac{(\Delta{}_{\rm mle}m_i - f_i)^2}{\sigma{}_{i}^2} + \frac{1}{2} \frac{f_i^2}{\sigma{}_{i}^2}\biggr)
\end{split}
\end{equation}

The output of the linear search is a set of two-dimensional arrays of: Depth, $\{\Delta{}_{i,j}\}_{i,j=1}^{I,J}$; depth variance, $\{\sigma_{\Delta_{i,j}}^2\}_{i,j=1}^{I,J}$; and likelihood ratio, $\{LR_{i,j}\}_{i,j=1}^{I,J}$. With $\{t_{i}\}_{i=1}^{I}$ and $\{D_{j}\}_{j=1}^{J}$ along the two primary array axes.

Figure \ref{fig:linear_output} shows some output from the linear search for an example light curve containing transits. The example light curve is astrophysically implausible, but \refnote{suitable} for demonstration purposes. The input transits have $5$~day durations and $1000$~ppm depth, and occur at $t={10,30,50}$, i.e. they correspond to a signal with a $20$~day period. The input model parameters are correctly recovered.

Since all resampled light curve array elements with no contributing observations have infinite uncertainties, they have no influence on either the depth or the likelihood ratio. These may include: null points padded to the beginning and end of the resampled light curve, which enable \textsc{cetra} to detect transits that straddle the beginning or end of the data without any special treatment within the linear search kernel; points within gaps in the observations, e.g. between observing cycles or during daytime (if the data are ground-based); and points between observations if the resampling frequency is higher than that of the observations. In some circumstances there may be no observations within the transit window, e.g. data gaps longer than any of the trial durations, or where $t<D/2$ due to the aforementioned padding. Output arrays elements are null in such circumstances, the latter case can be seen in panels (b) through (d) in Figure \ref{fig:linear_output}.

If required, the log-likelihood of the whole light curve for all transit models (i.e. the result of Equation \ref{eqn:logl}) can be calculated from the array of likelihood ratios described above. Simply add to them the log-likelihood of the light curve for a constant flux model (i.e. Equation \ref{eqn:logl} evaluated for $\Delta{}_{i,j}=0$), which is independent of $t$ and $D$ and hence need only be computed once for all pairs, at negligible expense.

% \begin{equation}\label{eqn:logl_flat}
%     \ln{}p(\mathbf{f}|t_s,D,0) = \sum_{i=0}^{N}\biggl(
%     -\frac{1}{2} \frac{f_i^2}{\sigma{}_{i}^2} - \frac{1}{2}\ln{(2\pi{}\sigma{}_{i}^2)}\biggr)
% \end{equation}

\subsection{Periodic search}
\label{sec:algo_periodic}

The periodic search proceeds as per \citet{DFM15}, albeit without allowing variation in the depths of transits contributing to a periodic signal.

In principle one might expect some low level of transit depth variation, in practice however, depth variation is likely to be negligible for the purposes of transit detection. Hence, \textsc{cetra} assumes that all transits contributing to a periodic signal have a common depth. If transit depths do vary, this will result in a reduction in the likelihood of the periodic signal. The likelihood function for the depth at each $t_i$,$D_j$ point is Gaussian, with depth $\Delta{}_{i,j}$, and variance $\sigma^2_{\Delta{}_{i,j}}$. The log-likelihood of the data for a model with some alternative depth, $Z$, is therefore:
\begin{equation}\label{eqn:mod_depth}
        \ln{p(\mathbf{f}|t_i,D_j,Z)} = 
        \ln{p(\mathbf{f}|t_i,D_j,\Delta{}_{i,j})}
        - \frac{1}{2} \frac{(\Delta{}_{i,j} - Z)^2}{\sigma^2_{\Delta_{i,j}}}
\end{equation}
And it follows that the likelihood ratio for this alternative model is:
\begin{equation}\label{eqn:mod_lrat}
        LR_{i,j}(Z) = LR_{i,j}
        - \frac{1}{2} \frac{(\Delta{}_{i,j} - Z)^2}{\sigma^2_{\Delta_{i,j}}}
\end{equation}

For any sequence of non-overlapping transits, the corresponding set of depths, depth variances, and likelihood ratios can be obtained from the results of the linear search. Nearest neighbour interpolation can be used to obtain approximates of these values where a transit falls between grid elements, an approach also adopted by \citet{DFM15}.
From these we can compute their maximum likelihood common depth and its variance, $Z_{i,j,k}$ and $\sigma{}^2_{Z_{i,j,k}}$ hereafter, which are their inverse variance weighted mean depth and the reciprocal of the sum of their weights.

%The set of transits in the linear space which contribute to a given periodic model are defined by a reference start time $t_i$, a transit duration $D_j$, and a period $P_k$. Reference start times greater than the period can be ignored, as the [$t_i$, $D_j$, $P_k$] model is a duplicate of the [$t_i+P_k$, $D_j$, $P_k$] model.

The sum of the results of Equation \ref{eqn:mod_lrat} for all transits contributing to a given periodic signal, having adopted $Z_{i,j,k}$, gives the likelihood ratio of the data given that periodic model versus a constant flux model ($LR_{i,j,k}$).
% I.e.:
% \begin{equation}\label{eqn:lr_periodic}
%         LR_{i,j,k} = \sum_{}^{}LR_{i,j}
%         - \frac{1}{2} \sum_{}^{}LR_{i,j}\frac{(\Delta{}_{i,j} - Z)^2}{\sigma^2_{\Delta_{i,j}}}
% \end{equation}
This follows logically, it is simply the log-likelihood of the data in all the transit windows for a (now multiple) transit model minus the same for a constant flux model, which itself is the likelihood ratio of the periodic signal model versus the constant flux model.

This method assumes that the linear search result contribution of each transit is independent of all others. \citet{DFM15} and \citet{nuance} note that this is only approximately true in their case, since they were fitting systematic trends in addition to transits. Their changing belief about the transits also changes their belief about the systematics model, thereby influencing all other transits. Since \textsc{cetra} requires a pre-whitened light curve, the assumption might be closer to reality, though due to the intricacies of detrending it is still unlikely to be strictly true. One point of note is that should the transit windows overlap then the results will be invalid, for this reason \textsc{cetra} requires that trial durations are smaller than the trial period.

For each trial period the two-dimensional array results that are the outputs of the linear search are essentially phase-folded, but with the additional common depth stipulation. For each, a smaller set of two dimensional arrays of depth, depth variance and likelihood ratios is produced. Copying each of these from the GPU back to the host machine would be a major bottleneck in the algorithm. However, the user is unlikely to be interested in anything other than the model with the maximum likelihood. Consequently, the periodic search kernel incorporates a reduction operation to identify the maximum likelihood element in the above arrays, and they return only this and the corresponding model parameters to the host machine.

\begin{figure}
  \begin{center}
    \includegraphics[width=0.45\textwidth,keepaspectratio]{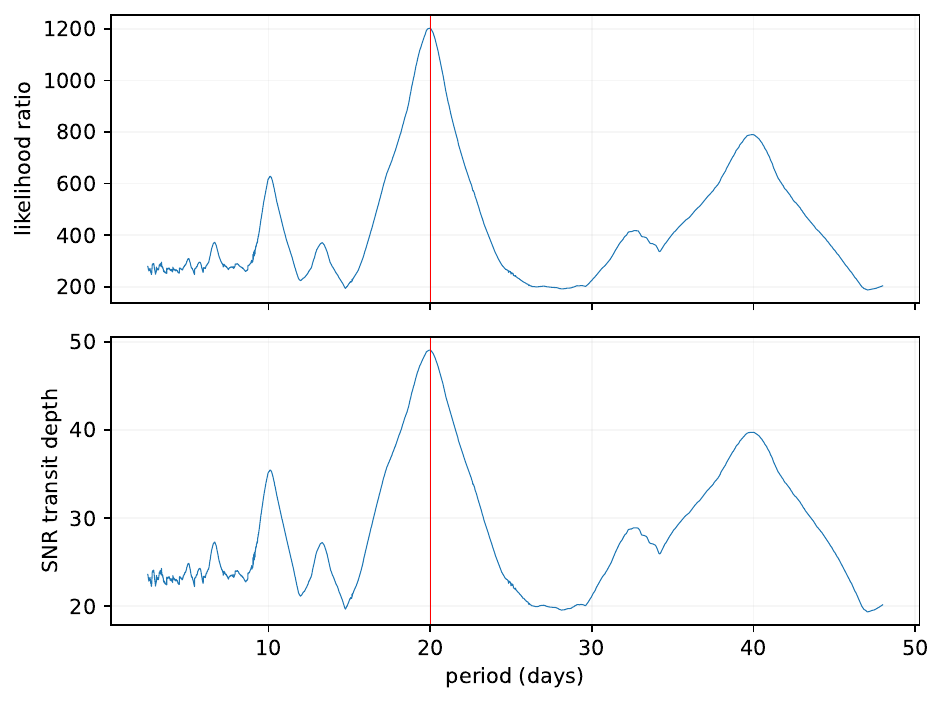}
    \caption{The result of the periodic signal search for the example light curve shown in Figure \ref{fig:linear_output}. The true period is $20$~days, indicated by the red line. The upper panel shows the likelihood ratio of the light curve for the maximum likelihood periodic transit model versus a model containing no transits. The lower panel shows the signal-to-noise ratio of the maximum likelihood periodic transit model. The unusually wide periodogram peak is caused by the astrophysically implausible light curve, which contains few transits with a large duration to period ratio. It is intended for algorithm demonstration purposes only. \refnote{The $\frac{1}{2}\times{}$ true period alias can be seen at 10 days, and the $2\times{}$ alias is seen at 40 days. Both are less prominent than the true peak}.}
    \label{fig:periodic_output}
  \end{center}
\end{figure}

The outcome of the periodic search is an array of periods with a corresponding array of their maximum likelihood ratios (demonstrated in the upper panel of Figure \ref{fig:periodic_output}), and their transit reference times, durations, depths and depth variances (demonstrated as a signal-to-noise ratio in the lower panel of Figure \ref{fig:periodic_output}). Should an array of log-likelihoods be required instead of likelihood ratios, then adding to them the log-likelihood of the constant flux model will achieve this. The log-likelihood of the constant flux model is itself a constant for a given detrended light curve, and this approach is common to the results of the linear search.

\section{Algorithm Performance}
\label{sec:performance}

\subsection{Simulated transit recovery}
\label{sec:performance_sim_scientific}

To evaluate the period recovery performance of \textsc{cetra} versus TLS, we generated 20\,000 synthetic transit light curves, scattered them using three different white noise levels, and checked to see which algorithm recovered the input periods. We did not include red noise in the light curves as the objective was to test detection algorithm performance, not detrending performance. Synthetic light curves were 173~days in length, sampled at 600~second cadence. Since real observational data regularly includes breaks in observations, particularly for long epoch baselines, we included a 7~day gap in the synthetic light curves starting at day~90. The three levels of white noise that we tested were 34, 160 and 800 ppm~hr$^{-1}$.

The injected transits had periods drawn randomly from a uniform distribution between $1.0$ and $86.5$~days, the upper bound meaning that at least two transits would always be present in the data unless one fell in the data gap. A transit duration was then drawn randomly from a uniform distribution with upper and lower bounds appropriate for the selected period (see Section \ref{sec:algo_grids}, above; and \citealt{TLS} figure 5). A transit depth was randomly selected from a log-uniform distribution bounded at 50 and 1\,000~ppm.
The injected transit model had the same impact parameter and limb darkening profile that TLS and \textsc{cetra} use by default.

We considered an input period to have been recovered if the output period was within 1~per~cent of the input period, or 1~per~cent of half or double the input period. An important detail is that both algorithms were always given exactly the same light curves, i.e. the parameters and noise properties were identical. The counts and percentages of recovered periods for each noise level and algorithm are shown in Table \ref{tab:wn_recovery_rates}, though the exact values of these are highly test-specific. Inferences about planet recovery rates from real surveys should not be drawn, since these are dependent on e.g. real transit parameter distributions, detrending performance, etc. As a consequence of this, and that these light curves contain only white noise and transits, and that detection thresholds are a choice for the user, we did not consider the level of contamination from false positives. The key takeaway is how the algorithms perform relative to each other.

\begin{table}
\centering
\caption{Results of TLS and \textsc{cetra} period recovery tests for 20\,000 synthetic light curves at three levels of white noise. Results for a subset of the 20\,000 is also shown, the 1\,165 samples which roughly correspond to Earth-Sun analogues. \refnote{The final column gives the relative gain in \textsc{cetra}'s recovery rate over the TLS recovery rate.} See text for details of the test parameters.}
\label{tab:wn_recovery_rates}
Entire set (20\,000 samples)
\begin{tabular}{|c|r|r|r|r|r}
\hline
  \multicolumn{1}{|c|}{error} &
  \multicolumn{2}{c|}{CETRA} &
  \multicolumn{2}{c|}{TLS}  &
  \multicolumn{1}{c|}{\% gain by} \\
  \multicolumn{1}{|c|}{ppm~hr$^{-1}$} &
  \multicolumn{1}{c|}{recovered} &
  \multicolumn{1}{c|}{\%}  &
  \multicolumn{1}{c|}{recovered}  &
  \multicolumn{1}{c|}{\%}  &
  \multicolumn{1}{c|}{CETRA} \\
\hline
 34    &  19\,726  &   98.6   & 19\,627 & 98.1  &   0.5 \\
 160   &  13\,268  &   66.3   & 12\,732 & 63.7  &   4.2 \\
 800   &   3\,290  &   16.5   &  2\,713 & 13.6  &  21.3 \\
\hline\end{tabular}
\vspace{0.2cm}\\
Earth analogue subset (1\,165 samples)
\begin{tabular}{|c|r|r|r|r|r}
\hline
  \multicolumn{1}{|c|}{error} &
  \multicolumn{2}{c|}{CETRA} &
  \multicolumn{2}{c|}{TLS}  &
  \multicolumn{1}{c|}{\% gain by} \\
  \multicolumn{1}{|c|}{ppm~hr$^{-1}$} &
  \multicolumn{1}{c|}{recovered} &
  \multicolumn{1}{c|}{\%}  &
  \multicolumn{1}{c|}{recovered}  &
  \multicolumn{1}{c|}{\%}  &
  \multicolumn{1}{c|}{CETRA} \\
\hline
 34    &  1\,132  &   97.2   & 1\,112 & 95.5  & 2  \\
 160   &      94  &    8.1   &     41 & 3.5   & 129  \\
 800   &       8  &    0.7   &      3 & 0.3   & 167  \\
\hline\end{tabular}
\end{table}

In this test \textsc{cetra} always outperforms TLS, but at low signal-to-noise ratios the improvement is more marked. For the test sample as a whole, in the $34$ and $160$~ppm~hr$^{-1}$ cases it correctly recovers up to $5$ per cent more periods, but in the $800$~ppm~hr$^{-1}$ case it recovers 21 per cent more. If we consider only the subset containing Earth-Sun analogues with transit depth of $75$ to $125$~ppm, and with two or three observed transits\footnote{In $173$~day simulated light curves, a period range of $58$ to $87$~days is roughly analogous to an Earth-like orbit in a two year light curve. The run time of TLS for two year light curves at 600~s cadence (see Section \ref{sec:performance_compute}) means that such a test would be unnecessarily expensive for this quick comparison.}, then \textsc{cetra} recovers $1.3\times{}$ and $1.7\times{}$ as many periods as TLS for the $160$ and $800$~ppm~hr$^{-1}$ samples, respectively. Earth-Sun analogues are highly prized, so improvements to the detection rate of such systems are valuable.

We attribute the improvement in sensitivity to the combination of a few factors:

Firstly, \citet{TLS} note (in their section 4.2) that TLS assumes a constant cadence in phase space. While the assumption is true on average, detection of longer period signals will be hindered to a greater degree. A significant cause of non-uniformity in phase-space cadence are gaps in the observational data. Gaps are almost always present in long epoch baseline light curves, as indeed they are in the above test. \textsc{Cetra} does not assume a constant cadence in phase (or time) space.

Secondly, TLS sweeps the light curve in phase space by the larger of one data sample or one per cent (by default) of the trial duration. This means that the true TLS $t_0$ grid spacing may be larger than expected for long period planets. For a two-transit periodic signal in a light curve, the effective $t_0$ grid spacing is half the observing cadence. This is potentially quite significant for long cadence (e.g. Kepler) data. \textsc{Cetra} always uses one per cent (by default) of the \textit{minimum} trial duration, it doesn't increase the effective grid spacing due to algorithmic constraints.

Finally, though this doesn't impact the above test as the flux uncertainties were uniform, TLS does not consider variable flux uncertainties when calculating the maximum likelihood depth of the transit model in a given window. TLS uses the arithmetic mean\footnote{via a cumulative sum optimisation} to determine the best model depth in the transit window, where \textsc{cetra} uses the inverse variance weighted mean.

\begin{figure}
  \begin{center}
    \includegraphics[width=0.45\textwidth,keepaspectratio]{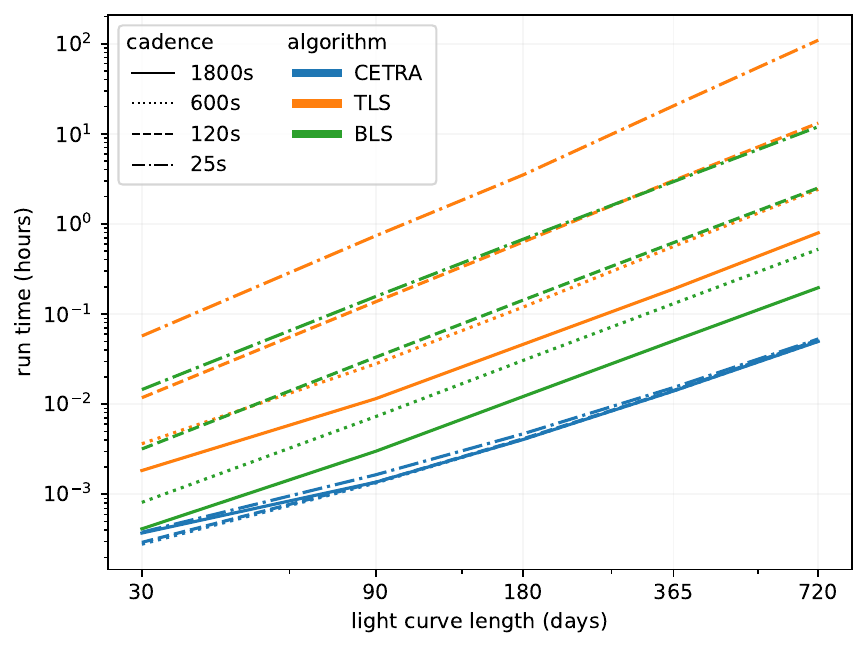}
    \caption{Compute times in CPU core hours for BLS and TLS, and GPU hours for \textsc{cetra}, for a range of light curve durations with various common cadences. TLS and BLS used a single core of an Intel~Xeon~Platinum~8358 CPU, while \textsc{cetra} used a single NVIDIA A100 GPU. See text for further discussion.}
    \label{fig:compute_time}
  \end{center}
\end{figure}

\begin{figure*}
  \begin{center}
    \includegraphics[width=\textwidth,keepaspectratio]{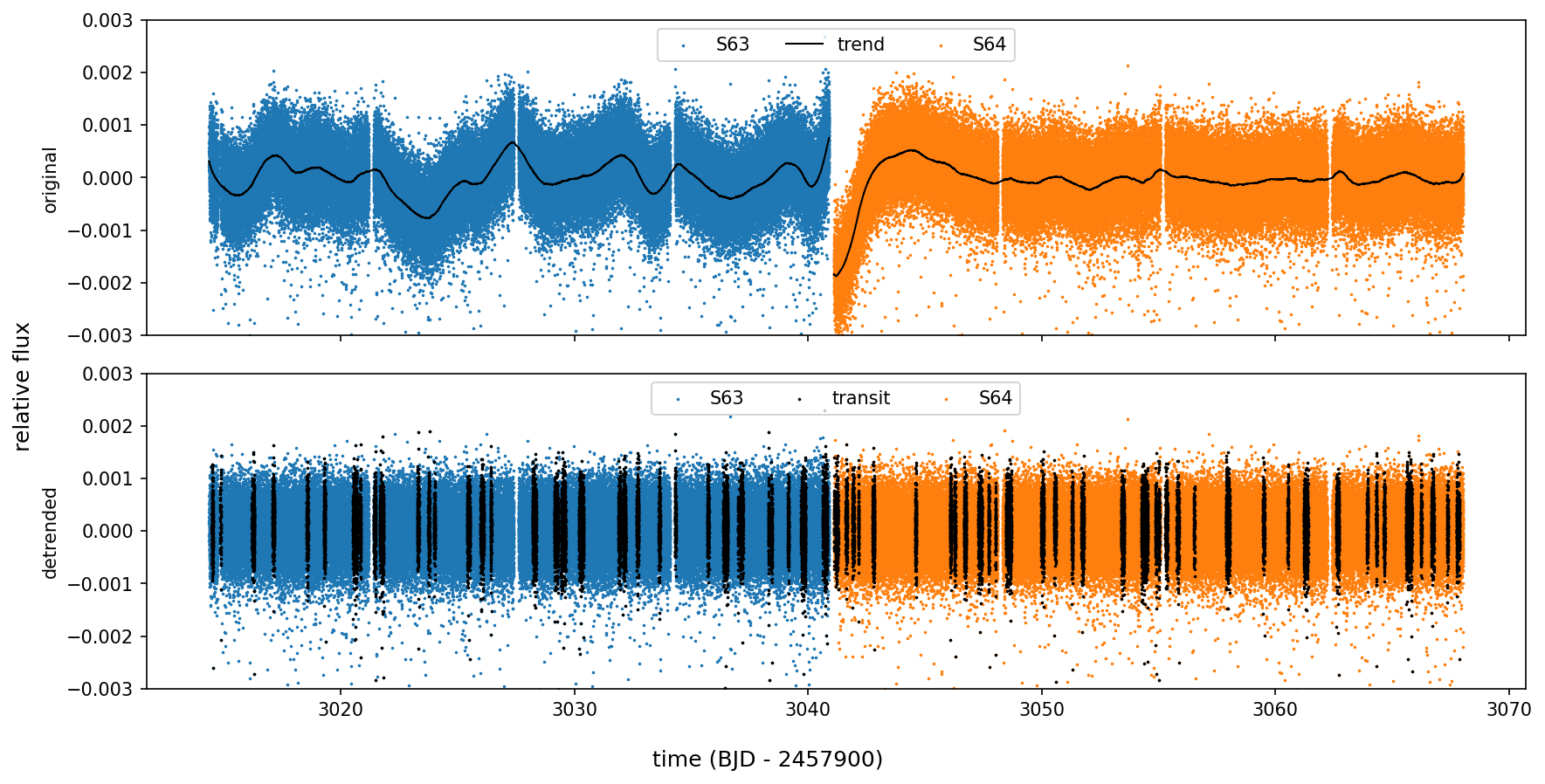}
    \caption{Original and detrended TESS PDCSAP flux of \mbox{HD 101581} at 20 second cadence. The locations of likely transits and the fitted trend are also shown.}
    \label{fig:HD101581_detrending}
  \end{center}
\end{figure*}

\subsection{Compute speed}
\label{sec:performance_compute}

We evaluated compute performance through comparison to BLS (via the \textsc{astropy} \textsc{timeseries} module which uses a C implementation via Cython) and TLS (which is Numba-compiled python code) for a range of light curve cadences and durations. The results are shown in Figure \ref{fig:compute_time}. Tested cadences were chosen to match existing and future data from the Kepler, TESS and PLATO missions. The 25~second cadence matches the PLATO short-cadence targets, and is comparable to the 20~second cadence of some TESS second extended mission targets. The 120~second cadence matches the majority of TESS targets. The 600~second cadence matches the PLATO long-cadence targets, and the 1800~second cadence matches that of Kepler long-cadence and K2 light curves.

The measured TLS and BLS compute times are for a single CPU core. We have typically found that performance per thread is reduced somewhat for TLS when using multiple threads internally, and the \textsc{astropy} implementation of BLS doesn't offer multithreading support. Parallelisation can also be achieved through running multiple separate, single-thread instances, each processing a different light curve or with different search parameters. Many system-architecture specific factors come into play when determining the optimal parallelisation method. To an approximate first order, a linear scaling with CPU core count can be applied, and consequently the compute times shown in Figure \ref{fig:compute_time} can be scaled to account for the user's own CPU core counts.

Computing overheads (resampling, grid generation, data transfers to/from GPU, model refinement, etc) were included for the \textsc{cetra}, TLS and BLS measurements where applicable. The \textsc{cetra} time includes both components of the search -- linear and periodic. BLS was run over the CETRA (/TLS) period and duration grids, skipping periods shorter than $5$~times the duration each time. All three algorithms were run using their default parameters.

The compute times shown in Figure \ref{fig:compute_time} were measured using a machine comprised of a pair of Intel~Xeon~Platinum~8358 CPUs, and a pair of NVIDIA~A100 GPUs, although BLS and TLS were run single-threaded and \textsc{cetra} used only one GPU. On a NVIDIA~RTX~A5000 GPU, which is very roughly one quarter of the monetary cost of an A100 at time of writing, \textsc{cetra} was only around one fifth slower.

\begin{figure*}
  \begin{center}
    \includegraphics[width=\textwidth,keepaspectratio]{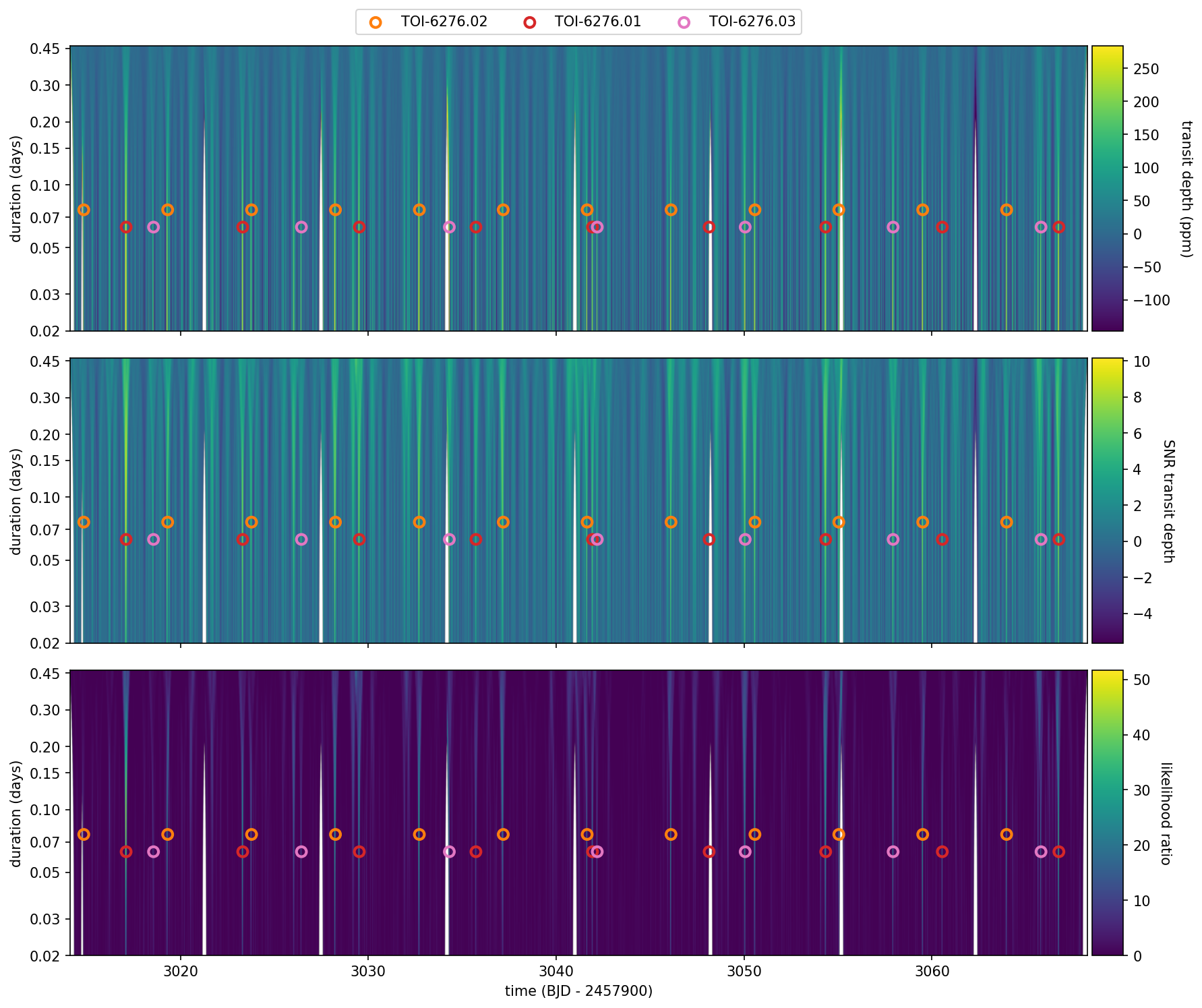}
    \caption{The output arrays of the linear transit signal search of \mbox{HD 101581}. This is the result of the first transit detection run, i.e. before any other signals have been removed. All panels have duration in days on the y-axis and $t_0$ on the x-axis. The upper panel shows the maximum likelihood transit depth in ppm, the middle panel shows the corresponding signal-to-noise ratio, and the lower panel shows the likelihood ratio of the transit model versus a model with no transit in that location. Gaps indicate regions where there are no data covering a transit window specified by the given $t_0$, duration pairing. \refnote{The locations of the transits identified in the subsequent periodic signal search are indicated (see Section \ref{fig:HD101581_periodic}).}}
    \label{fig:HD101581_linear}
  \end{center}
\end{figure*}

Figure \ref{fig:compute_time} demonstrates that \textsc{cetra} run times are largely independent of observing cadence, at least within the range of cadences that are likely to be useful for transit detection purposes. The time to complete the linear search is roughly a linear function of the number of light curve points, which is a function of light curve cadence and length\footnote{The duration grid size and start time grid density also play a role.}. The periodic search is only a function of light curve length, since this increases the highest checked period (assuming a requirement of 2 or more visible transits), and the number of transits that contribute to a signal at a given period. Importantly, the linear search is anywhere between a few times faster and a few orders of magnitude faster than the periodic search. For example, for the 25~second cadence, 30~day light curve the linear search took 0.26 seconds, while the periodic search took 0.94 seconds; and for the 1800~second cadence, 720~day light curve the linear search took 0.17~seconds, while the periodic search took 177.92~seconds. The periodic search times were essentially constant for a given light curve epoch baseline. Hence the cadence does not significantly impact the compute time. For this reason, resampling at a high frequency (see Section \ref{sec:algo_resample}) has a low cost.

\section{HD 101581 system recovery}
\label{sec:HD101581}

No demonstration of a transit detection algorithm would be complete without a test on real data. To this end, we attempted a re-detection of the \mbox{HD 101581} (\mbox{GJ 435}, \mbox{TOI-6276}, \mbox{TIC 397362481}) multi-planet system using \textsc{cetra}. \citet{kunimoto25} validated \mbox{HD 101581} b and c, a pair of Earth-sized planets found orbiting a K5V host. They also identify a third potential Earth-size planet, \mbox{TOI-6276.03}, but find the $7.9\sigma{}$ signal too weak to statistically validate. They used the TESS 120 second cadence Presearch Data Conditioning Simple Aperture Photometry \citep[PDCSAP;][]{Smith_2012, Stumpe_2012, Stumpe_2014} light curves from sectors 63 and 64, with the biweight detrending algorithm implemented in the \textsc{wotan} Python package \citep{wotan}, and TLS. We used the 20 second cadence PDCSAP light curve, with a Notch-like \citep{notch} detrending algorithm (details below), and \textsc{cetra}.

The following analysis is available as a Jupyter Notebook in the examples directory of the \textsc{cetra} GitHub repository.
%For detrending and detection we used a transit model generated with the \textsc{batman} package with the \textsc{cetra} default $b=0.32$, but with limb darkening parameters appropriate for a $T_{\rm eff}=4634 {\rm K}$, ${\rm log} g=4.71$, ${\rm [Fe}/{\rm H]}=-0.5$ star \citep[][table 3]{kunimoto25} based on https://exoctk.stsci.edu/limb_darkening
% But note that the limb darkening parameters are not too dissimilar to the cetra defaults, so tuning in this way made little difference to the detection significance, for this target

\subsection{Detrending}

Capitalising on the speed at which \textsc{cetra} is capable of checking for transit-like signals in linear light curve data, we developed a detrending algorithm inspired by the Notch filter algorithm of \citet{notch}. While not intended to be considered part of \textsc{cetra}, it is included in the code base for completeness. The algorithm traverses the light curve in linear space, searching for windows in which a transit-plus-quadratic model fits the observed flux better than a pure quadratic model based on some information criteria (Bayesian and Akaike are included). It searches over grids of $t_0$ and duration in exactly the same manner as the linear search described in Section \ref{sec:algo_linear}. With the above information, a basic model of the light curve is produced, which is then used to account for potential transits while refitting the baseline flux of the light curve. The algorithm also uses the non-transit regions of the model to measure a scaling factor to be applied to the flux uncertainties such that the standardised residuals resemble a unit Gaussian. In addition to those of the linear search (e.g. $t_0$ and duration grids), the main tuning parameters of the algorithm are the widths of the detection and flux fitting kernel windows, the type of information criterion used and its difference threshold.

We ran the above algorithm on the 20 second cadence \mbox{HD 101581} light curves from the two TESS sectors separately. We used transit detection and flux fitting kernel window widths of 1.5 and 1.0 days, respectively, and the Bayesian Information Criterion (BIC) with a difference threshold of 10. The input light curve, potential transit locations, fitted trend, and detrended light curve are shown in Figure \ref{fig:HD101581_detrending}. Detrending was performed using an NVIDIA RTX A5000 GPU and took 1 to 2 seconds per sector.

\subsection{Transit detection}

\begin{figure}
  \begin{center}
    \includegraphics[width=0.45\textwidth,keepaspectratio]{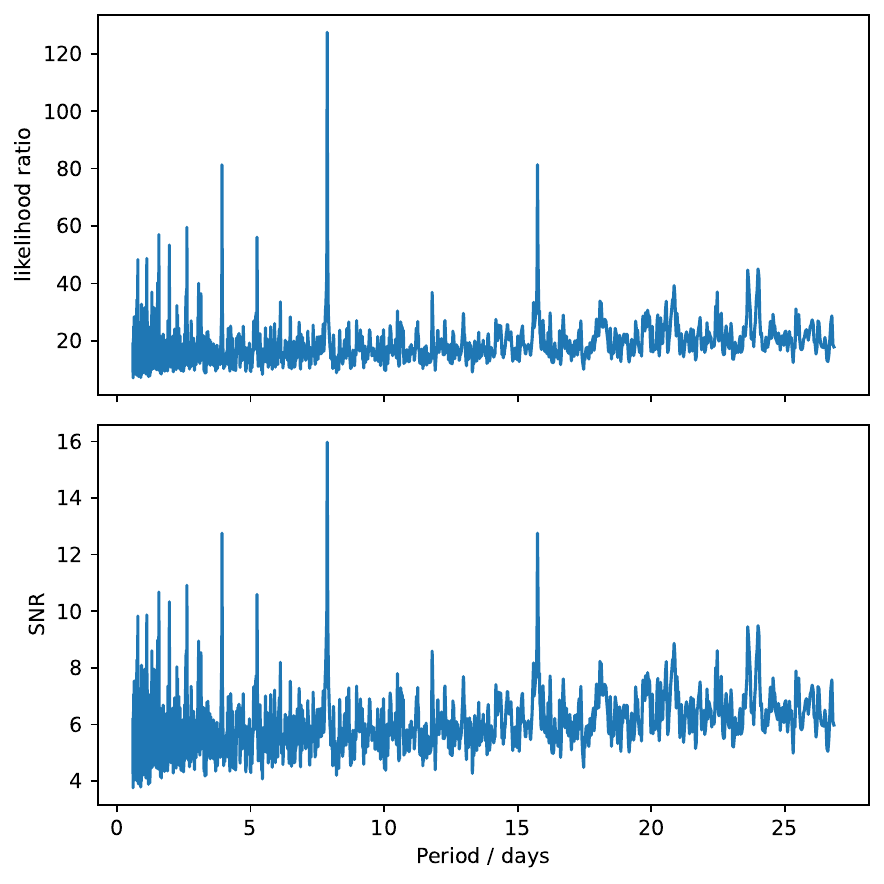}
    \caption{Periodograms of \mbox{HD 101581} after having masked the two most significant transit signals, which correspond to the planets \mbox{HD 101581} b and c. The likelihood ratio and signal-to-noise ratios are shown versus the trial period. The $16\sigma{}$ signal at 7.87 days corresponds to \mbox{TOI-6276.03}.}
    \label{fig:HD101581_periodic}
  \end{center}
\end{figure}

We ran the detrended \mbox{HD 101581} light curve through \textsc{cetra} with a maximum transit duration of 0.5 days, but otherwise used the default search parameters. We ran transit detection on the light curve three times, for each run after the first we nullified the flux within and around the highest likelihood transit signal from the previous search. For illustration purposes, the array output from the first run is shown in Figure \ref{fig:HD101581_linear}, and the periodogram from the final run is shown in Figure \ref{fig:HD101581_periodic}.

\begin{figure}
  \begin{center}
    \includegraphics[width=0.45\textwidth,keepaspectratio]{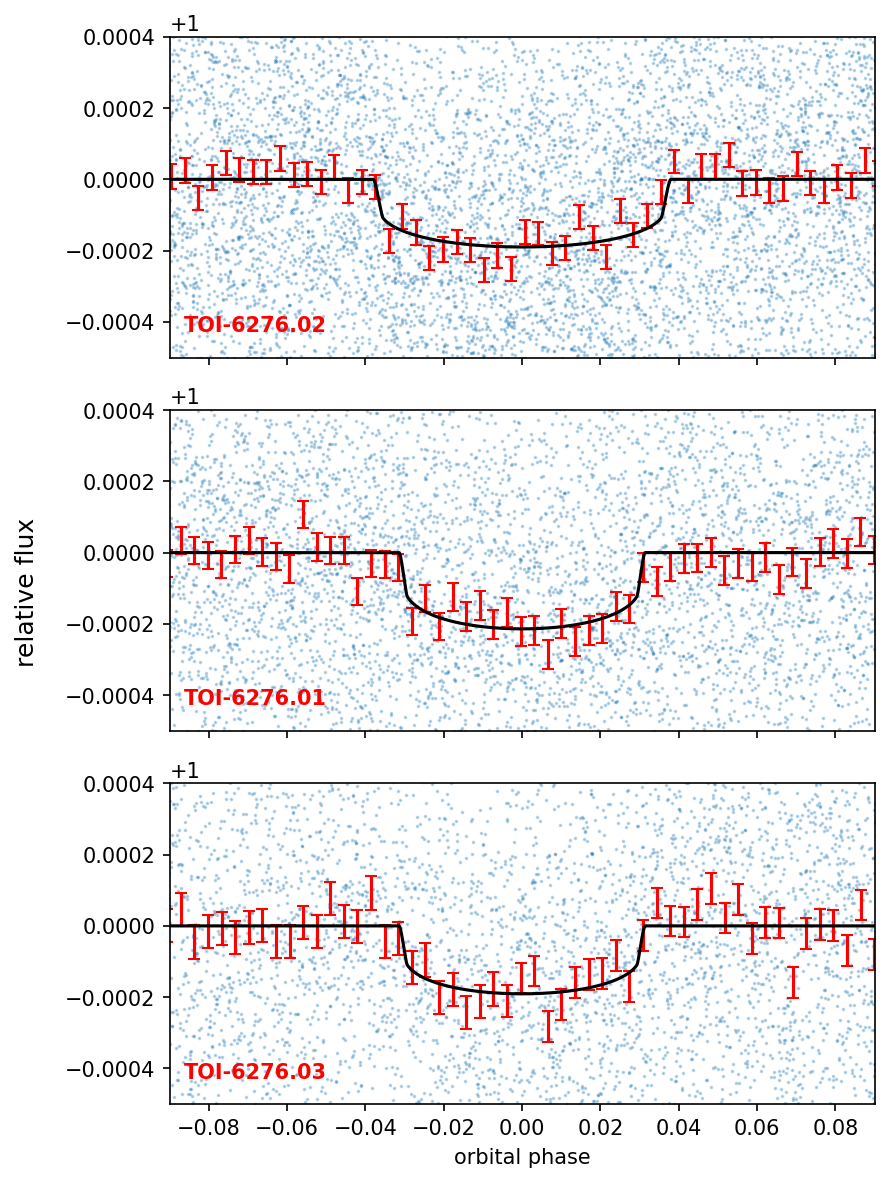}
    \caption{The three periodic transit signals identified by \textsc{cetra} in the TESS 20 second cadence PDCSAP light curve of \mbox{HD 101581}. Blue points are the detrended light curve, red error bars are the detrended light curve binned to 5 minute cadence, and the black line is the maximum likelihood transit model found by \textsc{cetra}.}
    \label{fig:HD101581_TCEs}
  \end{center}
\end{figure}

Details of the resultant periodic signals are given in Table \ref{tab:HD101581_detections} in the order of their detection. Figure \ref{fig:HD101581_TCEs} shows the maximum likelihood transit model over the data and binned data for each of the detected signals.

\begin{table}
\centering
\caption{The maximum likelihood transit parameters of the three transit detection passes on the TESS 20 second cadence PDCSAP light curve of \mbox{HD 101581}. The order is as-detected by \textsc{cetra}, from top to bottom. We report the SNR as the maximum likelihood transit depth divided by its uncertainty.}
\label{tab:HD101581_detections}
\begin{tabular}{|c|c|r|r|r|r}
\hline
  \multicolumn{1}{|c|}{TOI} &
  \multicolumn{1}{c|}{$t_0$} &
  \multicolumn{1}{c|}{period}  &
  \multicolumn{1}{c|}{duration}  &
  \multicolumn{1}{c|}{depth}  &
  \multicolumn{1}{c|}{SNR} \\
  \multicolumn{1}{|c|}{} &
  \multicolumn{1}{c|}{BJD-2457900} &
  \multicolumn{1}{c|}{days}  &
  \multicolumn{1}{c|}{days}  &
  \multicolumn{1}{c|}{ppm}  &
  \multicolumn{1}{c|}{} \\
\hline
 TOI-6276.02   &  3014.856   &    4.465   &  0.076   &  188  &  23.0 \\
 TOI-6276.01   &  3017.110   &    6.206   &  0.063   &  214  &  19.9 \\
 TOI-6276.03   &  3018.567   &    7.873   &  0.063   &  191  &  16.0 \\
\hline\end{tabular}
\end{table}

\subsection{Discussion}

A transit detection run on the above light curve with TLS v1.32 took of order fifteen minutes on a single thread of the test workstation. Parallelising over all six available CPU cores reduced the compute time to around four minutes. By contrast, a single run on one NVIDIA RTX A5000 GPU took 2.6 seconds, a factor 90 reduction in compute time versus TLS multithreaded. The obvious implication of this is that the user can process more targets in less time. It also enables the user to more cheaply sweep detrending and detection configurations in order to optimise searches for specific targets, and to run injection-recovery tests at moderate scales without the need for large cluster computing facilities.

Note that \textsc{cetra} did not detect \mbox{TOI-6276.03} in the 120 second cadence TESS light curves any better than \citet{kunimoto25} was able to with TLS. When run on the same detrended 20 second cadence light curves as \textsc{cetra}, TLS found all three signals with signal detection efficiency of around $18$ ($18.5$ for \mbox{TOI-6276.01}, $18.0$ for \mbox{TOI-6276.02}, and $17.8$ for \mbox{TOI-6276.03}). Possible reasons for the marginally improved sensitivity of \textsc{cetra} relative to TLS are given in Section \ref{sec:performance_sim_scientific}, but the signal of \mbox{TOI-6276.03} in the 120 second cadence light curve is not particularly affected by any of them. So \textsc{cetra} not being any more sensitive than TLS in this specific case is not surprising. It is not clear why the signal of \mbox{TOI-6276.03} is more significant in the short cadence light curve than the long. Possibly the detrending performance is better with a higher frequency of observations, we certainly found it easier to tune the detrending parameters for this data than for the long cadence data, but this is speculation. It is clear that there may be value in running detection algorithms on short cadence data rather than binning to longer cadence. This renders the compute efficiency improvements of \textsc{cetra} over existing algorithms that we demonstrated in Section \ref{sec:performance_compute} more important.

\citet{kunimoto25} note that the false positive probability of \mbox{TOI-6276.03} is low enough to satisfy the criteria for statistical validation when boosted due to its membership in a multi-planet system, though they did not consider it a statistically valid planet due to the low ${\rm SNR}=7.9$ in their analysis. We suspect the increase in the detection significance to ${\rm SNR}=16.0$ presented in this analysis would justify reconsideration of the statistical validity of the candidate.

\section{Summary}

\refnote{Motivated by the requirement to detect exoplanet transits in ever-increasing volumes of high cadence stellar light curves, and the apparent stalling of the trend toward increasing CPU clock speeds observed in the last few decades \citep{leiserson20},} we present a new transit detection algorithm \refnote{that takes advantage of GPUs and the highly parallel nature of the task. \textsc{Cetra}} has been implemented with CUDA for NVIDIA GPUs, although we're exploring the possibility of porting to other frameworks to enable processing on devices from other manufacturers. After preprocessing, it first evaluates the likelihood of a transit signal across grids of $t_0$ and transit duration in linear, i.e. time space. It then phase-folds the results of the linear search to generate a periodogram, from which periodic signals can be identified.

In common with the well established TLS code, \textsc{cetra} uses a true transit (or other) model in order to maximise detection sensitivity. When tested against 20\,000 synthetic light curves with varying noise levels, it was found to consistently outperform TLS, with a more marked improvement seen for low SNR planets, particularly those with long periods. The improvement is largely attributable to a more robust handling of gaps in the observational data, and overcoming of a $t_0$ sampling frequency limitation that specifically impacts detection of long period planets.

In addition to the improved sensitivity, through benchmarking against the most commonly used existing codes, BLS and TLS, \textsc{cetra} was demonstrated to be dramatically faster to run. The greatest improvement in compute speed of a few orders of magnitude is seen for high cadence light curves. Should the user wish to search only for monotransits, then \textsc{cetra} can do this in under a second even for high cadence light curves that are several years in length. Long-period planets appear as monotransits until sufficient epoch baseline and phase coverage is achieved by the observation campaign. A detection algorithm that is sensitive to them means decisions regarding key follow up observations can be made sooner.

\textsc{Cetra} detected the three Earth-sized planets in the \mbox{HD 101581} (\mbox{TOI-6276}) system within the TESS high cadence light curves. Two of the planets were previously detected with TLS and validated in the lower cadence light curves by \citet{kunimoto25}, but the third was found to be too low SNR to statistically validate. We found generally improved signal-to-noise ratios, with the signal of the previously unvalidated planet (\mbox{TOI-6276.03}) boosted from ${\rm SNR}=7.9$ to ${\rm SNR}=16.0$, which is potentially sufficient to reconsider the statistical validity of the planet.

The ability to rapidly search large datasets makes \textsc{cetra} ideally suited for current and future space-based exoplanet hunting missions (e.g. TESS, PLATO). The code is available from GitHub and PyPI under an open-source MIT license.

\section*{Acknowledgements}

We are grateful to the referee, Ren\'e Heller, for the careful reading and helpful comments. We acknowledge the support of the PLATO Mission Consortium, and from UKRI-STFC grants ST/X001628/1 and ST/X001571/1.

%%%%%%%%%%%%%%%%%%%%%%%%%%%%%%%%%%%%%%%%%%%%%%%%%%
\section*{Data Availability}

TESS light curves of \mbox{HD 101581} are publicly available online. Code and analyses are available from the \textsc{cetra} GitHub repository at \url{https://github.com/leigh2/cetra}.



\bibliographystyle{mnras}
\bibliography{refs} % if your bibtex file is called example.bib

\begin{thebibliography}{}
\makeatletter
\relax
\def\mn@urlcharsother{\let\do\@makeother \do\$\do\&\do\#\do\^\do\_\do\%\do\~}
\def\mn@doi{\begingroup\mn@urlcharsother \@ifnextchar [ {\mn@doi@} {\mn@doi@[]}}
\def\mn@doi@[#1]#2{\def\@tempa{#1}\ifx\@tempa\@empty \href {http://dx.doi.org/#2} {doi:#2}\else \href {http://dx.doi.org/#2} {#1}\fi \endgroup}
\def\mn@eprint#1#2{\mn@eprint@#1:#2::\@nil}
\def\mn@eprint@arXiv#1{\href {http://arxiv.org/abs/#1} {{\tt arXiv:#1}}}
\def\mn@eprint@dblp#1{\href {http://dblp.uni-trier.de/rec/bibtex/#1.xml} {dblp:#1}}
\def\mn@eprint@#1:#2:#3:#4\@nil{\def\@tempa {#1}\def\@tempb {#2}\def\@tempc {#3}\ifx \@tempc \@empty \let \@tempc \@tempb \let \@tempb \@tempa \fi \ifx \@tempb \@empty \def\@tempb {arXiv}\fi \@ifundefined {mn@eprint@\@tempb}{\@tempb:\@tempc}{\expandafter \expandafter \csname mn@eprint@\@tempb\endcsname \expandafter{\@tempc}}}

\bibitem[\protect\citeauthoryear{{Aigrain} \& {Irwin}}{{Aigrain} \& {Irwin}}{2004}]{ppp}
{Aigrain} S.,  {Irwin} M.,  2004, \mn@doi [\mnras] {10.1111/j.1365-2966.2004.07657.x}, \href {https://ui.adsabs.harvard.edu/abs/2004MNRAS.350..331A} {350, 331}

\bibitem[\protect\citeauthoryear{{Berta}, {Irwin}, {Charbonneau}, {Burke}  \& {Falco}}{{Berta} et~al.}{2012}]{Mearth_tda}
{Berta} Z.~K.,  {Irwin} J.,  {Charbonneau} D.,  {Burke} C.~J.,   {Falco} E.~E.,  2012, \mn@doi [\aj] {10.1088/0004-6256/144/5/145}, \href {https://ui.adsabs.harvard.edu/abs/2012AJ....144..145B} {144, 145}

\bibitem[\protect\citeauthoryear{{Borucki} et~al.,}{{Borucki} et~al.}{2010}]{kepler}
{Borucki} W.~J.,  et~al., 2010, \mn@doi [Science] {10.1126/science.1185402}, \href {https://ui.adsabs.harvard.edu/abs/2010Sci...327..977B} {327, 977}

\bibitem[\protect\citeauthoryear{Carter \& Agol}{Carter \& Agol}{2013}]{QATS}
Carter J.~A.,  Agol E.,  2013, \mn@doi [The Astrophysical Journal] {10.1088/0004-637X/765/2/132}, 765, 132

\bibitem[\protect\citeauthoryear{{Foreman-Mackey}, {Montet}, {Hogg}, {Morton}, {Wang}  \& {Sch{\"o}lkopf}}{{Foreman-Mackey} et~al.}{2015}]{DFM15}
{Foreman-Mackey} D.,  {Montet} B.~T.,  {Hogg} D.~W.,  {Morton} T.~D.,  {Wang} D.,   {Sch{\"o}lkopf} B.,  2015, \mn@doi [\apj] {10.1088/0004-637X/806/2/215}, \href {https://ui.adsabs.harvard.edu/abs/2015ApJ...806..215F} {806, 215}

\bibitem[\protect\citeauthoryear{{Garcia}, {Foreman-Mackey}, {Murray}, {Aigrain}, {Feliz}  \& {Pozuelos}}{{Garcia} et~al.}{2024}]{nuance}
{Garcia} L.~J.,  {Foreman-Mackey} D.,  {Murray} C.~A.,  {Aigrain} S.,  {Feliz} D.~L.,   {Pozuelos} F.~J.,  2024, \mn@doi [\aj] {10.3847/1538-3881/ad3cd6}, \href {https://ui.adsabs.harvard.edu/abs/2024AJ....167..284G} {167, 284}

\bibitem[\protect\citeauthoryear{{Hippke} \& {Heller}}{{Hippke} \& {Heller}}{2019}]{TLS}
{Hippke} M.,  {Heller} R.,  2019, \mn@doi [\aap] {10.1051/0004-6361/201834672}, \href {https://ui.adsabs.harvard.edu/abs/2019A&A...623A..39H} {623, A39}

\bibitem[\protect\citeauthoryear{{Hippke}, {David}, {Mulders}  \& {Heller}}{{Hippke} et~al.}{2019}]{wotan}
{Hippke} M.,  {David} T.~J.,  {Mulders} G.~D.,   {Heller} R.,  2019, \mn@doi [\aj] {10.3847/1538-3881/ab3984}, \href {https://ui.adsabs.harvard.edu/abs/2019AJ....158..143H} {158, 143}

\bibitem[\protect\citeauthoryear{{Howell} et~al.,}{{Howell} et~al.}{2014}]{K2}
{Howell} S.~B.,  et~al., 2014, \mn@doi [\pasp] {10.1086/676406}, \href {https://ui.adsabs.harvard.edu/abs/2014PASP..126..398H} {126, 398}

\bibitem[\protect\citeauthoryear{{Jenkins} et~al.,}{{Jenkins} et~al.}{2010}]{kepler_transits}
{Jenkins} J.~M.,  et~al., 2010, in {Radziwill} N.~M.,  {Bridger} A.,  eds,  Society of Photo-Optical Instrumentation Engineers (SPIE) Conference Series Vol. 7740, Software and Cyberinfrastructure for Astronomy. p. 77400D, \mn@doi{10.1117/12.856764}

\bibitem[\protect\citeauthoryear{{Kipping}}{{Kipping}}{2010}]{dont_bin}
{Kipping} D.~M.,  2010, \mn@doi [\mnras] {10.1111/j.1365-2966.2010.17242.x}, \href {https://ui.adsabs.harvard.edu/abs/2010MNRAS.408.1758K} {408, 1758}

\bibitem[\protect\citeauthoryear{{Kov{\'a}cs}, {Zucker}  \& {Mazeh}}{{Kov{\'a}cs} et~al.}{2002}]{BLS}
{Kov{\'a}cs} G.,  {Zucker} S.,   {Mazeh} T.,  2002, \mn@doi [\aap] {10.1051/0004-6361:20020802}, \href {https://ui.adsabs.harvard.edu/abs/2002A&A...391..369K} {391, 369}

\bibitem[\protect\citeauthoryear{{Kov{\'a}cs}, {Hartman}  \& {Bakos}}{{Kov{\'a}cs} et~al.}{2016}]{sim_detrend_doubt}
{Kov{\'a}cs} G.,  {Hartman} J.~D.,   {Bakos} G.~{\'A}.,  2016, \mn@doi [\aap] {10.1051/0004-6361/201527124}, \href {https://ui.adsabs.harvard.edu/abs/2016A&A...585A..57K} {585, A57}

\bibitem[\protect\citeauthoryear{{Kreidberg}}{{Kreidberg}}{2015}]{batman}
{Kreidberg} L.,  2015, \mn@doi [\pasp] {10.1086/683602}, \href {https://ui.adsabs.harvard.edu/abs/2015PASP..127.1161K} {127, 1161}

\bibitem[\protect\citeauthoryear{{Kunimoto} et~al.,}{{Kunimoto} et~al.}{2025}]{kunimoto25}
{Kunimoto} M.,  et~al., 2025, \mn@doi [\aj] {10.3847/1538-3881/ad9266}, \href {https://ui.adsabs.harvard.edu/abs/2025AJ....169...47K} {169, 47}

\bibitem[\protect\citeauthoryear{Leiserson, Thompson, Emer, Kuszmaul, Lampson, Sanchez  \& Schardl}{Leiserson et~al.}{2020}]{leiserson20}
Leiserson C.~E.,  Thompson N.~C.,  Emer J.~S.,  Kuszmaul B.~C.,  Lampson B.~W.,  Sanchez D.,   Schardl T.~B.,  2020, \mn@doi [Science] {10.1126/science.aam9744}, 368, eaam9744

\bibitem[\protect\citeauthoryear{{Leleu}, {Chatel}, {Udry}, {Alibert}, {Delisle}  \& {Mardling}}{{Leleu} et~al.}{2021}]{RIVERS}
{Leleu} A.,  {Chatel} G.,  {Udry} S.,  {Alibert} Y.,  {Delisle} J.~B.,   {Mardling} R.,  2021, \mn@doi [\aap] {10.1051/0004-6361/202141471}, \href {https://ui.adsabs.harvard.edu/abs/2021A&A...655A..66L} {655, A66}

\bibitem[\protect\citeauthoryear{{McQuillan}, {Aigrain}  \& {Roberts}}{{McQuillan} et~al.}{2012}]{kepler_variability}
{McQuillan} A.,  {Aigrain} S.,   {Roberts} S.,  2012, \mn@doi [\aap] {10.1051/0004-6361/201016148}, \href {https://ui.adsabs.harvard.edu/abs/2012A&A...539A.137M} {539, A137}

\bibitem[\protect\citeauthoryear{{Ofir}}{{Ofir}}{2014}]{ofir14}
{Ofir} A.,  2014, \mn@doi [\aap] {10.1051/0004-6361/201220860}, \href {https://ui.adsabs.harvard.edu/abs/2014A&A...561A.138O} {561, A138}

\bibitem[\protect\citeauthoryear{{Rauer} et~al.,}{{Rauer} et~al.}{2024}]{PLATO}
{Rauer} H.,  et~al., 2024, \mn@doi [arXiv e-prints] {10.48550/arXiv.2406.05447}, \href {https://ui.adsabs.harvard.edu/abs/2024arXiv240605447R} {p. arXiv:2406.05447}

\bibitem[\protect\citeauthoryear{{Ricker} et~al.,}{{Ricker} et~al.}{2015}]{TESS}
{Ricker} G.~R.,  et~al., 2015, \mn@doi [Journal of Astronomical Telescopes, Instruments, and Systems] {10.1117/1.JATIS.1.1.014003}, \href {https://ui.adsabs.harvard.edu/abs/2015JATIS...1a4003R} {1, 014003}

\bibitem[\protect\citeauthoryear{{Rizzuto}, {Mann}, {Vanderburg}, {Kraus}  \& {Covey}}{{Rizzuto} et~al.}{2017}]{notch}
{Rizzuto} A.~C.,  {Mann} A.~W.,  {Vanderburg} A.,  {Kraus} A.~L.,   {Covey} K.~R.,  2017, \mn@doi [\aj] {10.3847/1538-3881/aa9070}, \href {https://ui.adsabs.harvard.edu/abs/2017AJ....154..224R} {154, 224}

\bibitem[\protect\citeauthoryear{Smith et~al.,}{Smith et~al.}{2012}]{Smith_2012}
Smith J.~C.,  et~al., 2012, \mn@doi [Publications of the Astronomical Society of the Pacific] {10.1086/667697}, 124, 1000

\bibitem[\protect\citeauthoryear{Stumpe et~al.,}{Stumpe et~al.}{2012}]{Stumpe_2012}
Stumpe M.~C.,  et~al., 2012, \mn@doi [Publications of the Astronomical Society of the Pacific] {10.1086/667698}, 124, 985

\bibitem[\protect\citeauthoryear{Stumpe, Smith, Catanzarite, Van~Cleve, Jenkins, Twicken  \& Girouard}{Stumpe et~al.}{2014}]{Stumpe_2014}
Stumpe M.~C.,  Smith J.~C.,  Catanzarite J.~H.,  Van~Cleve J.~E.,  Jenkins J.~M.,  Twicken J.~D.,   Girouard F.~R.,  2014, \mn@doi [Publications of the Astronomical Society of the Pacific] {10.1086/674989}, 126, 100

\makeatother
\end{thebibliography}

%%%%%%%%%%%%%%%%%%%%%%%%%%%%%%%%%%%%%%%%%%%%%%%%%%

%%%%%%%%%%%%%%%%% APPENDICES %%%%%%%%%%%%%%%%%%%%%

% \appendix

% \section{Some extra material}

% If you want to present additional material which would interrupt the flow of the main paper,
% it can be placed in an Appendix which appears after the list of references.

%%%%%%%%%%%%%%%%%%%%%%%%%%%%%%%%%%%%%%%%%%%%%%%%%%

% Don't change these lines
\bsp	% typesetting comment
\label{lastpage}
\end{document}